\documentclass[twocolumn]{aastex62}

\usepackage{epstopdf}
\usepackage{threeparttable}

\received{ 01/02/2018}
\revised{ 13/02/2018}
\accepted{20/02/2018 }
\submitjournal{ApJL}

\shorttitle{On the chemical abundance of Miras in clusters}
\shortauthors{D'Orazi et al.}

\begin{document}

\title{On the chemical abundances of Miras in clusters: V1 in the metal-rich globular NGC~5927\footnote{Based on spectra collected with the WINERED spectrograph available as a visitor instrument at the ESO New Technology Telescope (NTT), La Silla, Chile (ESO Proposal: 098.D-0878(A), PI: G.~Bono)}}

\correspondingauthor{Valentina D'Orazi}
\email{valentina.dorazi@oapd.inaf.it}

\author{V. D'Orazi}
\affiliation{INAF Osservatorio Astronomico di Padova,  vicolo dell'Osservatorio 5, 35122, Padova, Italy}
\affiliation{Monash Centre for Astrophysics, School of Physics and Astronomy, Monash University, Melbourne, VIC 3800, Australia}

\author{D. Magurno}
\affiliation{University of Roma Tor Vergata, via della Ricerca Scientifica 1, 00133, Roma, Italy}
\affiliation{INAF Osservatorio Astronomico di Roma, via Frascati 33, 00040, Monte Porzio Catone RM, Italy}

\author{G. Bono}
\affiliation{University of Roma Tor Vergata, via della Ricerca Scientifica 1, 00133, Roma, Italy}
\affiliation{INAF Osservatorio Astronomico di Roma, via Frascati 33, 00040, Monte Porzio Catone RM, Italy}

\author{N. Matsunaga}
\affiliation{Department of Astronomy, The University of Tokyo, 7-3-1 Hongo, Bunkyo-ku, Tokyo 113-0033, Japan}
\affiliation{Laboratory of infrared High-resolution spectroscopy (LiH), Koyama Astronomical Observatory, Kyoto-Sangyo University, ~Motoyama, Kamigamo, Kita-ku, Kyoto 606-8555, Japan}


\author{V.F. Braga}
\affiliation{Instituto Milenio de Astrofisica, Av. Vicu\~na Mackenna 4860, 782-0436 Macul, Santiago, Chile}
\affiliation{Departamento de Fisica, Facultad de Ciencias Exactas, Universidad Andres Bello, Av. Fernandez Concha 700, Las Condes, Santiago, Chile}

\author{S.S. Elgueta}
\affiliation{Department of Astronomy, The University of Tokyo, 7-3-1 Hongo, Bunkyo-ku, Tokyo 113-0033, Japan}

\author{K. Fukue}
\affiliation{Laboratory of infrared High-resolution spectroscopy (LiH), Koyama Astronomical Observatory, Kyoto-Sangyo University, ~Motoyama, Kamigamo, Kita-ku, Kyoto 606-8555, Japan}

\author{S. Hamano}
\affiliation{Laboratory of infrared High-resolution spectroscopy (LiH), Koyama Astronomical Observatory, Kyoto-Sangyo University, ~Motoyama, Kamigamo, Kita-ku, Kyoto 606-8555, Japan}

\author{L. Inno}
\affiliation{Max-Planck Institute for Astronomy, D-69117, Heidelberg, Germany}

\author{N. Kobayashi}
\affiliation{Kiso Observatory, Institute of Astronomy, The University of Tokyo, 10762-30 Mitake, Kiso-machi, Kiso-gun, Nagano 397-0101, Japan}
\affiliation{Laboratory of infrared High-resolution spectroscopy (LiH), Koyama Astronomical Observatory, Kyoto-Sangyo University, ~Motoyama, Kamigamo, Kita-ku, Kyoto 606-8555, Japan}

\author{S. Kondo}
\affiliation{Laboratory of infrared High-resolution spectroscopy (LiH), Koyama Astronomical Observatory, Kyoto-Sangyo University, ~Motoyama, Kamigamo, Kita-ku, Kyoto 606-8555, Japan}

\author{M. Monelli}
\affiliation{Instituto de Astrofísica de Canarias, Calle Via Lactea s/n, E38205 La Laguna, Tenerife, Spain}

\author{M. Nonino}
\affiliation{INAF Osservatorio Astronomico di Trieste, Via G. B. Tiepolo 11, 34143 Trieste, Italy}

\author{N. Przybilla}
\affiliation{Institut f\"ur Astro- und Teilchenphysik, Universit\"at Innsbruck, Technikerstrasse 25, 6020, Innsbruck, Austria}

\author{H. Sameshima}
\affiliation{Laboratory of infrared High-resolution spectroscopy (LiH), Koyama Astronomical Observatory, Kyoto-Sangyo University, ~Motoyama, Kamigamo, Kita-ku, Kyoto 606-8555, Japan}

\author{I. Saviane}
\affiliation{European Southern Observatory, Karl-Schwarzschild-Str. 2, 85748 Garching bei Munchen, Germany}

\author{D. Taniguchi}
\affiliation{Department of Astronomy, The University of Tokyo, 7-3-1 Hongo, Bunkyo-ku, Tokyo 113-0033, Japan}

\author{F. Thevenin}
\affiliation{Universit\'e de La C\^ote d'Azur, OCA, Laboratoire Lagrange CNRS, BP. 4229, 06304, Nice Cedex, France}

\author{M. Urbaneja-Perez}
\affiliation{Institut f\"ur Astro- und Teilchenphysik, Universit\"at Innsbruck, Technikerstrasse 25, 6020, Innsbruck, Austria}

\author{A. Watase}
\affiliation{Division of Science, Graduate School, Kyoto Sangyo University, Motoyama, Kamigamo, Kita-ku, Kyoto 606-8555, Japan}


\author{A. Arai}
\affiliation{Laboratory of infrared High-resolution spectroscopy (LiH), Koyama Astronomical Observatory, Kyoto-Sangyo University, ~Motoyama, Kamigamo, Kita-ku, Kyoto 606-8555, Japan}

\author{M. Bergemann}
\affiliation{Max-Planck Institute for Astronomy, D-69117, Heidelberg, Germany}

\author{R. Buonanno}
\affiliation{University of Roma Tor Vergata, via della Ricerca Scientifica 1, 00133, Roma, Italy}
\affiliation{INAF Osservatorio Astronomico d'Abruzzo, Via Mentore Maggini snc, Loc. Collurania, 64100 Teramo, Italy}

\author{M. Dall'Ora}
\affiliation{INAF Osservatorio Astronomico di Capodimonte, Salita Moiariello 16, 80131 Napoli, Italy}

\author{R. Da Silva} 
\affiliation{SSDC, via del Politecnico snc, 00133 Roma, Italy}
\affiliation{INAF Osservatorio Astronomico di Roma, via Frascati 33, 00040, Monte Porzio Catone RM, Italy}

\author{M. Fabrizio}
\affiliation{SSDC, via del Politecnico snc, 00133 Roma, Italy}
\affiliation{INAF Osservatorio Astronomico di Roma, via Frascati 33, 00040, Monte Porzio Catone RM, Italy}

\author{I. Ferraro}
\affiliation{INAF Osservatorio Astronomico di Roma, via Frascati 33, 00040, Monte Porzio Catone RM, Italy}

\author{G. Fiorentino}
\affiliation{INAF Osservatorio Astronomico di Bologna, Via Ranzani 1, 40127 Bologna, Italy}

\author{P. Francois}
\affiliation{Observatoire de Paris, PSL Research University, CNRS, Place Jules Janssen, 92190 Meudon, France}

\author{R. Gilmozzi}
\affiliation{European Southern Observatory, Karl-Schwarzschild-Str. 2, 85748 Garching bei Munchen, Germany}

\author{G. Iannicola}
\affiliation{INAF Osservatorio Astronomico di Roma, via Frascati 33, 00040, Monte Porzio Catone RM, Italy}

\author{Y. Ikeda}
\affiliation{Photocoding, 460-102 Iwakura-Nakamachi, Sakyo-ku, Kyoto 606-0025, Japan}
\affiliation{Laboratory of infrared High-resolution spectroscopy (LiH), Koyama Astronomical Observatory, Kyoto-Sangyo University, ~Motoyama, Kamigamo, Kita-ku, Kyoto 606-8555, Japan}

\author{M. Jian}
\affiliation{Department of Astronomy, The University of Tokyo, 7-3-1 Hongo, Bunkyo-ku, Tokyo 113-0033, Japan}

\author{H. Kawakita}
\affiliation{Laboratory of infrared High-resolution spectroscopy (LiH), Koyama Astronomical Observatory, Kyoto-Sangyo University, ~Motoyama, Kamigamo, Kita-ku, Kyoto 606-8555, Japan}
\affiliation{Department of Astrophysics and Atmospheric Sciences, Faculty of Science, Kyoto Sangyo University, Motoyama, Kamigamo, Kita-ku, Kyoto 603-8555, Japan}

\author{R.P. Kudritzki}
\affiliation{Institute for Astronomy, University of Hawaii, 2680 Woodlawn Drive, Honolulu, HI 96822, USA}

\author{B. Lemasle}
\affiliation{Astronomisches Rechen-Institut, Zentrum f\"ur Astronomie der Universit\"at Heidelberg, M\"onchhofstr. 12-14, 69120, Heidelberg, Germany}

\author{M. Marengo}
\affiliation{Department of Physics and Astronomy, Iowa State University, Ames, IA 50011, USA}

\author{S. Marinoni}
\affiliation{SSDC, via del Politecnico snc, 00133 Roma, Italy}
\affiliation{INAF Osservatorio Astronomico di Roma, via Frascati 33, 00040, Monte Porzio Catone RM, Italy}

\author{C. E. Mart\'inez-V\'azquez}
\affiliation{Cerro Tololo Inter-American Observatory, National Optical Astronomy Observatory, Casilla 603, La Serena, Chile}

\author{D. Minniti}
\affiliation{Instituto Milenio de Astrofisica, Av. Vicu\~na Mackenna 4860, 782-0436 Macul, Santiago, Chile}
\affiliation{Departamento de Fisica, Facultad de Ciencias Exactas, Universidad Andres Bello, Av. Fernandez Concha 700, Las Condes, Santiago, Chile}

\author{J. Neeley}
\affiliation{University of Florida, Department of Astronomy, 211 Bryant Space Science Center  P. O. 112055, Gainesville, FL, USA}

\author{S. Otsubo}
\affiliation{Division of Science, Graduate School, Kyoto Sangyo University, Motoyama, Kamigamo, Kita-ku, Kyoto 606-8555, Japan}

\author{J.L. Prieto}
\affiliation{N\'ucleo de Astronom\'ia, Facultad de Ingenier\'ia y Ciencias, Universidad Diego Portales, Ej\'ercito 441, Santiago, Chile}

\author{B. Proxauf}
\affiliation{Max Planck Institute for Solar System Research Justus-von-Liebig-Weg 3 37077 G\"ottingen}

\author{M. Romaniello} 
\affiliation{European Southern Observatory, Karl-Schwarzschild-Str. 2, 85748 Garching bei Munchen, Germany}

\author{N. Sanna}
\affiliation{INAF Osservatorio Astronomico di Arcetri, Largo Enrico Fermi 5, 50125 Firenze, Italy}

\author{C. Sneden}
\affiliation{Department of Astronomy C1400, 1 University Station, University of Texas, Austin, TX 78712, USA}

\author{K. Takenaka}
\affiliation{Division of Science, Graduate School, Kyoto Sangyo University, Motoyama, Kamigamo, Kita-ku, Kyoto 606-8555, Japan}

\author{T. Tsujimoto}
\affiliation{National Astronomical Observatory of Japan, 2-21-1 Osawa, Mitaka, Tokyo 181-8588}

\author{E. Valenti}
\affiliation{European Southern Observatory, Karl-Schwarzschild-Str. 2, 85748 Garching bei Munchen, Germany}

\author{C. Yasui}
\affiliation{National Astronomical Observatory of Japan, 2-21-1 Osawa, Mitaka, Tokyo 181-8588}

\author{T. Yoshikawa} 
\affiliation{Edechs, 17-203 Iwakura-Minami-Osagi-cho, Sakyo-ku, Kyoto 606-0003, Japan}

\author{M. Zoccali}
\affiliation{Instituto Milenio de Astrofisica, Av. Vicu\~na Mackenna 4860, 782-0436 Macul, Santiago, Chile}
\affiliation{Pontificia Universidad Catolica de Chile, Instituto de Astrofisica, Av. Vicu\~na Mackenna 4860, Santiago, Chile}



\begin{abstract}
We present the first spectroscopic abundance determination of
iron, $\alpha$-elements (Si, Ca and Ti) and sodium for the Mira variable V1 in the metal-rich globular
cluster
NGC~5927. We use high-resolution ($R\sim 28,000$), high signal-to-noise ratio ({$\sim$}200) spectra
collected with WINERED, a near-infrared (NIR) spectrograph covering simultaneously
the wavelength range 0.91--1.35~$\mu$m.
The effective temperature and the surface gravity
at the pulsation phase
of the spectroscopic observation
were estimated using both optical ($V$) and NIR
time-series photometric
data. We found that the Mira is metal-rich ([Fe/H]$=-0.55 \pm 0.15$) and
moderately $\alpha$-enhanced ([$\alpha$/Fe]$=0.15 \pm 0.01$, $\sigma$=0.2). 
 These values
agree quite well with the mean cluster abundances based on high-resolution optical
spectra of several cluster red giants available in the literature
([Fe/H]$=-0.47 \pm 0.06$, [$\alpha$/Fe]$=+0.24 \pm 0.05$).
We also found a Na abundance of $+0.35 \pm 0.20$ that is higher than
the mean cluster abundance based on optical
spectra ($+0.18 \pm 0.13$). However, the lack of similar spectra for cluster red giants and that
of corrections for departures from local-thermodynamical equilibrium prevents us from establishing
whether the difference is intrinsic or connected with multiple populations.
These findings indicate a strong similarity between optical and NIR metallicity
scales in spite of the difference in the experimental equipment, data analysis and in the adopted spectroscopic diagnostics. 
\end{abstract}



\section{Introduction}

Radial variables have several key 
advantages compared with static stars, making them good stellar tracers. 
They can be easily identified even in crowded stellar fields using differential 
photometry. They are typically good distance indicators, 
 and
individual distances can be estimated with an accuracy better than a few percent.  
Classical Cepheids and Miras do provide the unique opportunity to estimate 
individual ages, since their periods are 
anti-correlated with their individual ages.  
This implies the opportunity to trace radial gradients across the main Galactic components 
(thin disk: \citealt{dasilva16};   bulge: \citealt{kunder13,zoccali16};   halo: \citealt{fiorentino15a})
and in nearby stellar systems \citep{martinezvazquez16b}.  

Miras play a crucial role in this context, since their parent population covers 
a broad range in stellar ages: from a few hundred Myr up to the age of 
globular clusters (GCs). This means they are ubiquitous, because they are present
in intermediate-age to old
stellar environments.
We are interested in cluster Miras, 
since they allow us to have 
  {\it a priori} robust
information
concerning the chemical composition, the environment and the evolutionary channel where 
they come from. Moreover, they allow us to develop a homogeneous metallicity scale between 
  Miras and other stars in GCs, mainly red giants,
widely investigated \citep[][and references therein]{carretta09}. We focused
our attention on V1 in NGC~5927, since this is a well 
known metal-rich GC \citep{pancino10}. 
It should be noted that V1 is listed as a irregular variable (Lb class)
 in \cite{clement01}, but we identified this object as a Mira, or
 an intermediate type between a Mira and a semi-regular variable,
 according to its periodic variation with a large infrared amplitude
 (Fig. 1 in \citealt{sloan10}, and see also Section~\ref{sec:parameters}).
 Its ampliude, {$\sim$}0.4~mag, is around the lower end of
 the infrared amplitudes of Miras (e.g., \citealt{matsunaga09}).
 As illustrated in Fig. 9 of Sloan et al. (2010), V1 lies on the period-luminosity
 relation of Miras (and relatively large-amplitude semi-regulars).
The selection of a metal-rich GC
was mainly driven by the fact that the occurrence of Miras appears to be 
correlated with iron abundance \citep{frogel98}. 
The reasons why we decided to collect NIR high-resolution,
high signal-to-noise ratio spectra with WINERED are manifold:
a) we are mainly interested in Miras located in the bulge (field, globulars);
this means stellar environments that are crowded and heavily reddened.
b) WINERED covers a substantial wavelength range (0.91--1.35 $\mu$m) and it
is characterized by a high spectral resolution ($R \sim 28,000$, WIDE mode).
Miras are late-type stars, this means that they are intrinsically brighter in the 
quoted wavelength range.
Thus, we have the opportunity to 
identify many iron and $\alpha$--element lines. Moreover, WINERED is also characterized 
by a very high-sensitivity and impressive throughputs ---from $\sim$30\% in the $z$-band to more than 50\% in the $J$-band--- 
when compared with similar NIR spectrographs \citep{ikeda16}. 
c) WINERED can also collect spectra with very high spectral resolution ($R \sim 68,000$, HIRES mode, \citealt{otsubo16}), 
covering either the $Y$ or the $J$ band. 

We present in this {\it Letter} the first spectroscopic characterisation of a cluster Mira
done by using a high-resolution near-infrared 
  spectrum ($z$, $Y$, $J$ bands), and report its
abundances for iron, $\alpha$-elements and sodium.  

\section{Observations and data reduction}

We observed the Mira V1 in NGC~5927 with the WIDE mode, $R \sim 28,000$,
of  WINERED, a PI instrument attached to the 3.58-m New Technology Telescope (NTT)
at La Silla observatory, ESO, Chile. The observation was done at around
08:25 on 2017 Feb 13 (UT), and the weather condition was fairly stable.
We obtained two integrations for the target of 300 seconds each,
and the co--added spectrum is expected to give a S/N higher than 200.
The spatial spread function shows
a FWHM of about 1.4~arcsec including
the seeing and the tracking accuracy. 
Two integrations were done
with the target at different positions within the slit (i.e.\  
AB positions).

The reduction was performed by using the automated pipeline developed  
by the WINERED team
\citep[see e.g.][]{taniguchi18}.
This pipeline produces continuum-normalized spectra after
standard analysis steps including 
bad pixel masking, sky subtraction, flat-fielding, scattered
light subtraction, spectrum extraction, wavelength 
calibration and continuum normalization. 
We used ThAr lamp data for the wavelength calibration and
the wavelengths were corrected to the standard air scale.

\subsection{Tellurics subtraction}
\begin{figure*}
	\includegraphics[width=\textwidth]{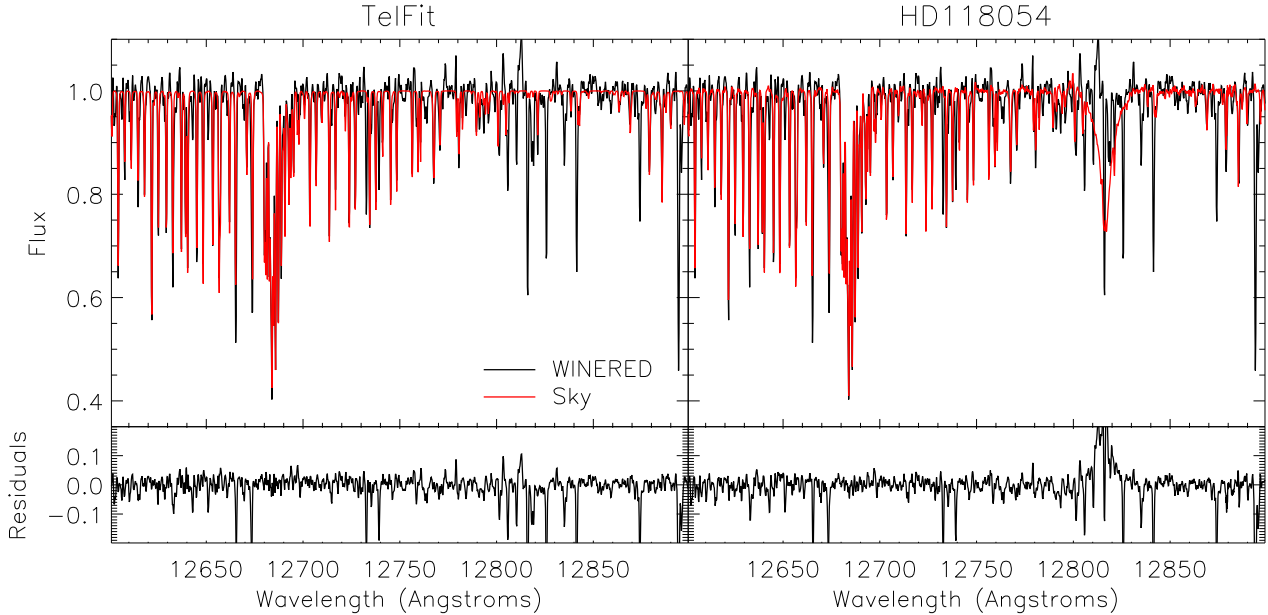}
    \caption{Left column: comparison between the original WINERED spectrum of
  the Mira V1 in NGC~5927 (black line) and the synthetic sky modelled with {\sc TelFit} (red line).  
Residuals are shown in the bottom panel. Right column: as before, but 
  the spectrum of the standard telluric star HD~118054 was used instead of {\sc TelFit}. Note that the residuals show intrinsic lines of the target.}
    \label{fig:telfit}
\end{figure*}

The main spurious features affecting every stellar spectrum are caused by 
the Earth's atmosphere. Molecular absorption bands are observed at fixed and well known 
wavelengths, but their strength depends on the current atmospheric conditions. 
In particular, NIR bands are more affected by tellurics than the optical bands. 
These lines are removed from the raw spectrum before performing any kind of 
abundance analysis, to avoid possible mis-identification and systematics in the 
estimate of the equivalent widths. The most common approach relies on the use 
of telluric standard stars. An early-type star with few and weak metallic lines 
is observed, close in time and in airmass to the target star, and its spectrum is subtracted 
from the target
  \citep[][and references therein]{sameshima18}.
This technique faces three main problems: a) atmospheric conditions 
can change rapidly during the night, thus it is not trivial to observe a telluric 
standard close in time and in sky position to the individual targets; 
b) it requires a significant investment in telescope time; 
c) telluric lines and stellar photospheric lines might be blended, thus limiting the accuracy of the correction \citep{sameshima18}.
We decided to adopt a different approach and to use the synthetic sky modeller {\sc TelFit} by \citet{gullikson14} 
to compute
  the telluric spectra for individual target spectra.
The synthetic sky was modelled independently 
for the 20 spectral orders of WINERED ($\Delta \lambda \simeq 300$~\AA). 
This approach allows us to properly trace the variation in spectral resolution when moving from the 
blue ($\lambda \simeq 9,200$~\AA, $R \sim 28,000$) to the red ($\lambda \simeq 13,400$~\AA, 
$R \sim 30,000$) regime of WINERED (see Fig. 5 in \citealt{ikeda16}).
A comparison between {\sc TelFit} 
and the standard telluric approach is shown in Fig.~\ref{fig:telfit} 
for the range 12,600--12,900~\AA. The subtraction of tellurics based on synthetic sky spectra 
and on the standard star agree quite well, and indeed both the residuals are of 
the order of 3\%. 
However, note that the standard star shows a disturbing hydrogen absorption feature at 12,818 \AA,
which is completely absent with the synthetic sky approach, compromising the identification of some 
useful absorption lines (see Fig. \ref{fig:synth1}).
The approach based on synthetic sky spectra appears very promising, since the spectrum 
of the telluric standard was collected in ideal conditions, i.e.\  26~min after the Mira 
spectrum and with a minimal difference in airmass (1.04 vs 1.19).

\section{Results and discussions}

\subsection{Stellar atmospheric parameters}\label{sec:parameters}
As a first step, we 
derived the radial velocity (RV) 
of our target through cross-correlation with a grid of synthetic spectra in selected wavelength 
regions, from 11,700 to 13,000~\AA.  We 
determined a heliocentric velocity of RV= $-$105.2 $\pm$ 2.0 km s$^{-1}$ (based on 34 spectral lines), which agrees 
quite well with the cluster average value given by the Harris's catalogue 
(\citeyear{harris96}, 2010 update\footnote{\url {http://physwww.mcmaster.ca/~harris/mwgc.dat}}) 
of $-107.5 \pm 0.9$~km~s$^{-1}$ and by \citet{simmerer13b} of $-104.03 \pm 5.03$~km~s$^{-1}$.
Note that the velocity amplitude of the Miras minimally affects this finding, since their 
typical variation is 
  {$\sim$}10~km~s$^{-1}$
\citep{wood79}.   
Since our spectral coverage does not grant the inclusion of a sufficient 
number of Fe~{\sc i} and (most crucially) Fe~{\sc ii} lines, 
the atmospheric parameters were adopted from photometric properties.
More specifically, effective temperature ($T_{\rm eff}$) 
  was
obtained using $J-K_{\rm s}$ colours and the calibration by \citet{alonso99}, assuming the reddening value 
from \citet{harris96} of $E(B-V)=0.45$, which 
  corresponds to
$E(J-K_{\rm s})=0.23$ based on the extinction law of \citet{cardelli89}. In order to estimate the pulsation phase, we used the ASAS-SN light curve \citep{shappee14,kochanek17} 
which covers the epoch of our spectroscopic observation well (330 phase points
  from 2016 Mar to 2017 Jul; period of the Mira P=202 days from \citet{sloan10}).
Although the angular resolution of the ASAS-SN is low (15~arcsec)
for our target in the GC, its light curve clearly
indicates that the target was near a minimum 
and the $V$-band magnitude is estimated at $15.3 \pm 0.1$~mag.
Unfortunately, we have no recent infrared photometry for the target,
and thus we 
  used
a light curve obtained at 1.4-m Infrared Survey Facility
about ten years ago. \citet{matsunaga06b} obtained 49 photometric
points which show periodic variation from 2002 Mar to 2005 Aug
with an amplitude of {$\sim$}0.4~mag in $K_{\rm s}$.  
Assuming that the phase lag between $V$-band and $K$-band light curves is
0.0--0.2 (with $V$ preceding, see e.g. \citealt{smith06}),
the $K$-band phase for the spectroscopic data is 0.3--0.1~cycles
before the minimum leading to $J-K_{\rm s}=1.3\pm 0.05$~mag and $K_{\rm s}=8.9 \pm 0.15$~mag from the IRSF light curve.
$V-K_{\rm s}$ is then $6.4\pm 0.2$~mag, which corresponds to $(V-K_{\rm s})_0=5.1\pm 0.4$~mag, while $(J-K_{\rm s})_0=1.05\pm 0.05$~mag,
with the reddening corrected. 

We obtained a $T_{\rm eff}=3600$~K using the $J-K_{\rm s}$ colours and 
3500~K using the $V-K_{\rm s}$ colours and the calibration by 
\citet{bessell98}. We adopted the former one, since the NIR photometry 
was collected simultaneously. The $J-K_{\rm s}$ is also less prone 
to reddening uncertainties when compared with $V-K_{\rm s}$ colour, 
since E(J-K)/E(V-K)=0.19 mag \citep{cardelli89}.  An error of 
100~K is thus a plausible uncertainty.
We also applied the temperature scale based on the reddening-free method of 
line-depth ratios constructed by \citet{taniguchi18}.
Some lines of their 81 line pairs
cannot be measured in the crowded spectrum of V1 in NGC~5927,
however, we estimated $T_{\rm eff} =3665 \pm 63$~K. 
The current value is consistent with the estimate based on the colour--temperature transformations,
thus suggesting that they are minimally affected by a possible reddening variation and/or dust formation in warm Miras.
Note that this temperature estimate was slightly extrapolated, 
since the temperatures of the calibrating stars used by \citet{taniguchi18} range from
3780 to 5400~K.  

From the photometric $T_{\rm eff}$, assuming a mass of $M=0.6~M_\odot$,
a true distance modulus of $\mu=14.44$~mag (\citealt{harris96}), and the bolometric correction for $K$ magnitudes 
by \citet{buzzoni10}, we
  estimated the surface
gravity of $\log g=0.0\pm 0.2$,
where the error comprises contributions from all the different sources of uncertainty
(i.e.\  temperature, luminosity).
A microturbulence of $\xi =2.0$~km~s$^{-1}$ was set, following prescriptions from the literature for this kind of cool, giant stars \citep[e.g.][]{origlia13}; note also that 
\citet{nowotny10} and \citet{lebzelter14} imposed a value of $\xi=2.5$~km~s$^{-1}$ for Miras, 
in agreement, within the errors, with our value (see Table~\ref{tab:abundances}).

\subsection{Abundance analysis}

The determination of elemental abundances was
carried out via spectral synthesis calculations using the driver
{\it synth} in {\sc moog} by C.~Sneden (\citeyear{sneden73}, 2017 version) and
the MARCS grid of spherical model atmospheres \citep{gustafsson08},
with $\alpha$ enhancements. The above mentioned atmospheric parameters 
  were
adopted, along with a global metallicity in the model atmosphere of [M/H]$=-0.5$\footnote{We adopt the standard notation for abundances, whereby [X/H]$=A({\rm X})-A({\rm X})_\odot$ and $A({\rm X})=\log (N_{\rm X}/N_{\rm H})+12$.} (see Harris's catalogue). The following crucial step includes the building of the line list. We 
carefully selected only atomic lines that are proven to be relatively isolated, unblended 
and not affected by departures from
local thermodynamical equilibrium (LTE). Our spectrum encompasses several K~{\sc i} lines, but we 
discarded this species since non-LTE corrections are not available for the lines under scrutiny (i.e.\  $\lambda = 11,772.838, 12,432.27$ and 12,522.134~\AA). 
Moreover, we only kept lines that provide abundances for the Sun ($T_{\rm eff}$=5770 K, logg=4.44, $\xi$=0.9 km~s$^{-1}$, [M/H]=0, \citealt{dorazi17}), and Arcturus 
($T_{\rm eff}$=4286 K, logg=1.66, $\xi$=1.74 km s$^{-1}$, [M/H]=$-$0.52, \citealt{ramirez11}) in compliance with literature values: 
all our measurements are in agreement within 0.1~dex with \citet{asplund09} and \citet{ramirez11}, respectively.
Our choice, though limiting the number of lines and species that can be measured, allows us to infer reliable abundance measurements, with no major systematics affecting our values. 
Our final line list includes Na~{\sc i}, Fe~{\sc i}, Si~{\sc i}, Ca~{\sc i} 
and Ti~{\sc i} lines and is given in Table~\ref{tab:linelist}, where we report for each line the atomic parameters, i.e.\  excitation potential and  
$\log gf$. The latter come from different literature sources, including values by Kurucz{\footnote{\url{http://kurucz.harvard.edu/linelists.html}} 
and the most recent computations for Ti~{\sc i} lines by \citet{lawler13}.
In order to perform the comparison between observed and synthetic spectrum, 
we have selected six wavelength regions with each interval covering {$\sim$}200~\AA:
this means synthetic calculations for more than 1000~\AA, by covering all the spectral lines of interest. 
An example of a spectral region that we have selected for our chemical analysis is shown in Fig.~\ref{fig:synth1}, whereas a zoom on the Ti line at 12,671~\AA\ is displayed in Fig.~\ref{fig:zoom}.
Our target has a low effective temperature and to properly locate the continuum
we included molecular line lists for CH, CN, CO and OH from B.~Plez (private communication). 
The determination of C, N, O abundances is a tricky task because of their inter-dependency and because they are changing during the star's evolution.
To add further complications, since our star is a GC member, 
all the three elements under discussion are involved in the hot hydrogen burning that is commonly accepted to happen in a fraction of the cluster first generation stars 
(the so-called {\it multiple population scenario}, see \citealt{gratton12} for an extensive review).
Moreover, the WINERED spectral coverage does not grant the inclusion of the CO bandhead and/or OH features located in the $H$- and $K$-band, 
which are commonly used to derive abundances for carbon and oxygen. Conversely, our spectra are populated with a large number of CN features. 
Thus, it is not straightforward to get insights on the initial content for C, N, O and on the amount 
of depletion/enhancement that has occurred as the star gets evolved.
For the current purpose we computed a grid of different synthetic spectra
assuming different CNO abundances, and finding the best fit that minimises 
the $\chi^2$. Note that this approach does not allow us to derive C, N and O 
abundances, since different combination can provide similar $\chi^2$ values.
We are taking into account these molecular features to improve the continuum 
determination.
\begin{figure*}
	\includegraphics[width=\textwidth,height=0.45\textheight]{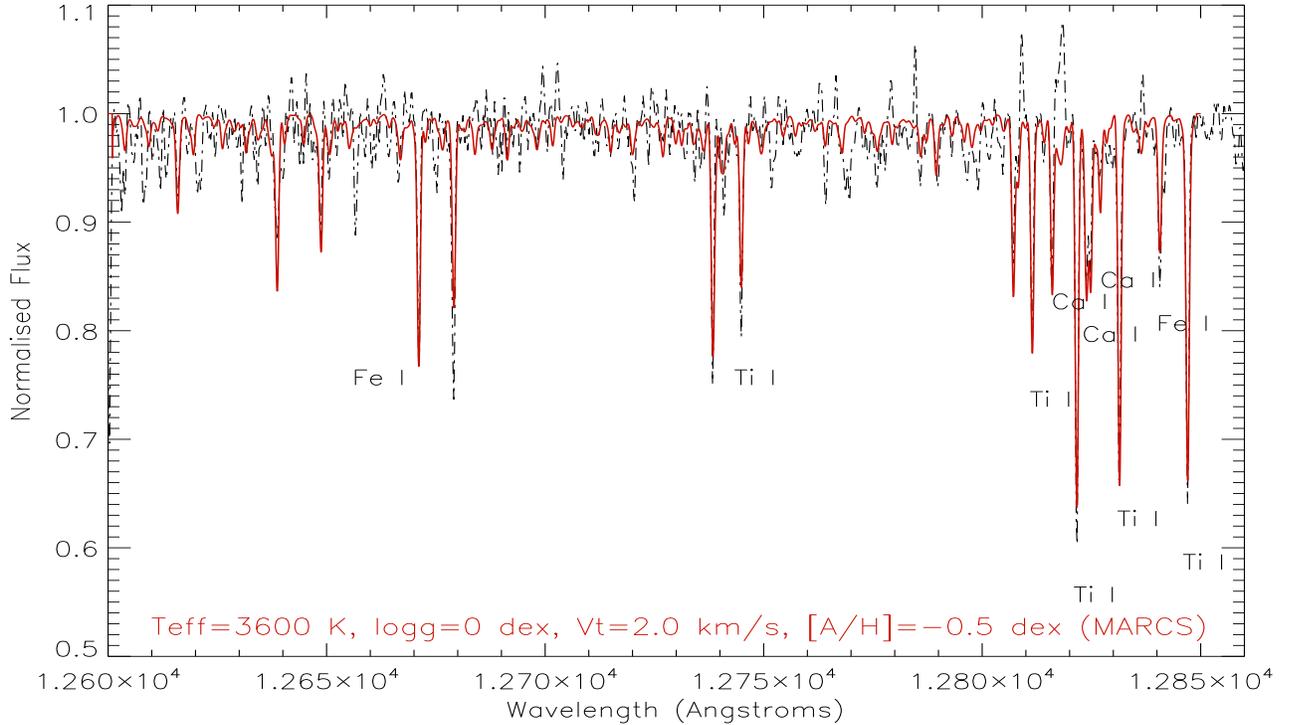}
     \caption{Example of a spectral window exploited to compare synthetic (solid line) and observed (dot-dashed line) spectrum.
     Key diagnostics for abundances are marked (Fe, Ca and Ti)}.
     \label{fig:synth1}
\end{figure*}
\begin{figure}
\includegraphics[width=0.5\textwidth]{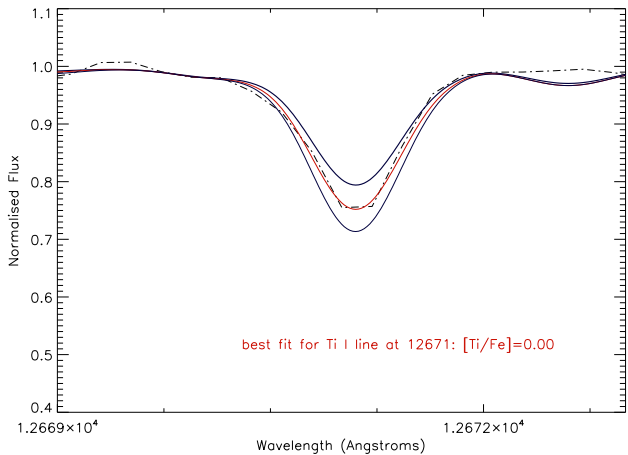}
\caption{Zoom on the Ti~{\sc i} line at 12,671~\AA. Different spectral syntheses (solid lines) are for  [Ti/Fe]$=0.00\pm 0.2$, compared with the observed spectrum (dot-dashed line).}
    \label{fig:zoom}
  \end{figure}

Chemical abundances are affected by internal uncertainties due to two main sources of error: 
(1) uncertainties on the best fit determination (that takes into account continuum displacement and line measurements) and (2) errors related to the adopted set of stellar parameters. For the first kind of error we adopted 
the standard deviation (r.m.s) from the mean abundances as given from different spectral lines: typical values are in the range 0.07--0.10~dex. 
To estimate errors due to stellar parameters ($T_{\rm eff}$, $\log g$, $\xi$ and [M/H]) we have proceeded in the standard way, that is changing 
  each parameter one by one
and evaluating the corresponding variation on the resulting abundances. 
Thus, temperature, gravity, microturbulence and global metallicity were
changed by $\pm 100$~K, $\pm 0.2$~dex, $\pm 0.5$~km~s$^{-1}$ and $\pm 0.1$~dex; we found errors on the [X/Fe] ratios of 0.10--0.12~dex.
We then added in quadrature the four different error contributions and calculated
the final error related to best fit and stellar parameters as:
\begin{equation}\label{eq:sensi}                                        
\sigma= \sqrt{\sigma^2_{\rm best} +                                         
\sigma_{T_{\rm eff}}^2 +                                                                  
 \sigma_{\log g}^2 +\sigma_{\xi}^2+
\sigma_{\rm [M/H]}^2},
\end{equation}                                                          
  see the result given in Table~\ref{tab:abundances}.
We note that, given the very good agreement for benchmark stars such as  the Sun and Arcturus, major systematic uncertainties 
should not affect our abundance values at a level larger than {$\sim$}0.1~dex.

\subsection{Results and concluding remarks}

Our results are given in Table~\ref{tab:abundances}, along with the corresponding total uncertainty (best fit procedure and errors due to stellar parameters).
The metallicity, [Fe/H]$=-0.55 \pm 0.15$, 
is in good agreement, within the observational uncertainties, with previous determinations from optical spectroscopy of GC giant members. \cite{harris96} gives for NGC 5927 a value of [Fe/H]$=-0.49$, whereas
\citet{pancino17} found a slightly larger metal content, 
[Fe/H]$=-0.39\pm 0.04$. 
Very recently, \citet{mura18} have presented high-resolution, FLAMES/UVES observations for a sample of seven red giants in this cluster, 
reporting a mean metallicity of [Fe/H]$=-0.47\pm 0.02$ (error of the mean, with r.m.s.$=$0.06~dex).
Concerning the other chemical species,  the cluster is included in the Gaia ESO survey but \citet{pancino17} have published abundances only for Mg and Al (see their Table~2).

On the other hand, \citet{mura18} have derived abundances for iron-peak, $\alpha$
and heavy elements (e.g.\  Ba  and Eu). In the last column of Table~\ref{tab:abundances} 
we show their values of [X/Fe] ratios for the species in common with the present study.
The two sets of elemental abundances agree quite well. 
Titanium and silicon abundances are slightly higher in \citet{mura18}, but still 
compatible within the uncertainties, whereas there is an excellent agreement 
between the two Ca measurements. The current findings suggest a 
modest $\alpha$-enhancement, less than {$\sim$}0.2~dex.
RGB stars in the Bulge display a steady decrease in $\alpha$-enhancement as a function of the iron content 
\citep{gonzalez11} approaching solar abundances ([$\alpha$/Fe]$\sim$0) in the 
metal-rich regime ([Fe/H]$\ge 0$). The trend for GCs ---targets that are old ($t \ge 10$~Gyr) 
and almost coeval--- for iron abundances higher than $-0.7$~dex is not well 
established, due to their paucity and for the limited number that has been 
spectroscopically investigated \citep{zoccali16}. However, the current 
estimate suggests that NGC~5927 is located in the lower envelope of the $\alpha$-enhancements
typical of   GCs \citep{pritzl05b,mura18}.  

\begin{table}
\centering
\caption{Stellar parameters ($T_{\rm eff}, \log g$ and $\xi$) and abundances for our target star. The corresponding uncertainties are given (see text for details).The
last column gives the cluster average abundances along with the standard deviation by \citet{mura18}}. \label{tab:abundances}
\small
\begin{tabular}{lrc} 
\hline\hline
   &   Mira V1  &  Cluster average\\
\hline
$\alpha$               & $15^h28^m15^s.2$  & --- \\
$\delta$            & -50\arcdeg 38\arcmin 09\arcmin \arcmin .0 &  ---  \\
$K_{\rm s}^a$  (mag)     &  $8.9 \pm 0.15$ & --- \\
$A K_{\rm s}^a$   (mag)    & $\sim$0.4 &  --- \\
P$^b$ (days)       &  202   &   --- \\
$T_{\rm eff}$   (K)           &  $3600  \pm 100$   & --- \\
$\log g$  (cgs)              &  $0.00  \pm 0.20$  & --- \\
$\xi$           (km~s$^{-1}$) &  $2.0   \pm 0.5$   & --- \\
$[$Fe/H$]$               &  $-0.55 \pm 0.15$  & $-0.47 \pm 0.06$ \\
$[$Na/Fe$]^{c}$          &  $+0.35 \pm 0.20$  & $+0.18 \pm 0.13$ \\
$[$Si/Fe$]$              &  $+0.14 \pm 0.15$  & $+0.24 \pm 0.08$ \\
$[$Ca/Fe$]$             &  $+0.13 \pm 0.20$  & $+0.15 \pm 0.04$ \\
$[$Ti/Fe$]$              &  $+0.17 \pm 0.13$  & $+0.32 \pm 0.06$ \\
\hline\hline
\end{tabular}
\begin{tablenotes}
\item $^a$~Mean $K_{\rm s}$-band magnitude and amplitude \citep{matsunaga06b}
\item $^b$~\cite{sloan10}
\item $^c$~Element affected by p-capture reactions.
\end{tablenotes}
\end{table}

As for Na, we obtained [Na/Fe]$=0.35 \pm 0.20$, to be compared with $0.18 \pm 0.13$ of the cluster average. 
The sodium content deserves a specific discussion.
There is a debate in the literature as to whether second-generation (i.e.\  Na-rich) AGB stars do exist \citep[see e.g.][]{campbell13,lapenna15,wang16,maclean16}. 
The Na abundances reported by \citet{mura18} are corrected for departures from LTE following prescriptions given in the 
INSPECT database\footnote{\url{http://inspect.coolstars19.com/index.php?n=Main.HomePage}} that are not available for our Na~{\sc i} line at 12,679~\AA.
Thus, this could in principle explain part of the discrepancy with our value; however, there is another critical point that has to be considered. 
Na is one of the species involved in proton-capture reaction processes that occur in GCs. All the GCs with a sufficient number of stars analysed display internal variation in Na \citep[e.g.][]{gratton12}. 
In particular, while the first generation stars have Na in agreement with field stars (at the corresponding metallicity), the second-generation GC stars exhibit a significant Na enhancement. 
At the present stage, we cannot confirm (or disprove) that Mira V1 in NGC 5927 belongs to the cluster second generation, because
of the low precision and also because we lack a control sample of red giants acquired with the same instrument.

The abundance analysis of Mira stars has been
affected by a number of long-standing problems: incompleteness of atomic and molecular line list \citep{uttenthaler15}, 
inhomogeneous atmospheres and complex circumstellar 
envelops \citep{hron15}, together with nonlinear phenomena in the cool molecular region located 
between the photosphere and the expanding molecular shell. These issues and the impact of both 
hydrostatic and dynamical models have been addressed in detail by \citet{lebzelter15}.  
  These difficulties are at least partly reduced because
we are dealing with a Mira that is on average warmer than typical Miras. 
The interesting finding in the current approach is the similarity 
between optical and NIR abundance scale in spite of the difference in the adopted
spectroscopic diagnostics. However, a more quantitative analysis of the impact of 1D versus
3D and static versus dynamical atmosphere models \citep{chiavassa18} would be
highly desirable in view of the unprecedented opportunity to observe Mira stars in Local Volume galaxies with the next generation of ELTs \citep{bono17}.

\begin{table}
\centering
\caption{Atomic line list for Na~{\sc i} (Species$=$11.0), 
Si~{\sc i} (14.0), Ca~{\sc i} (20.0), Ti~{\sc i} (22.0) and Fe~{\sc i} (26.0)
lines that we used for the abundance analysis.The [X/H] ratios are given in the last column.}
\label{tab:linelist}
\small
\begin{tabular}{lcccr}
\hline\hline
Wavelength & Species & E.P.  & $\log gf$ & [X/H]\\
(\AA)      &         & (eV)  &           \\
\hline
12,679.144  & 11.0    & 3.614 & $-$0.04 &  $-$0.20\\
12,679.144  & 11.0    & 3.614 & $-$1.34 &     ---\\
12,679.224  & 11.0    & 3.614 & $-$2.65&    ---\\
11,984.198  & 14.0    & 4.926 & $+$0.19 &  $-$0.55\\
11,991.568  & 14.0    & 4.916 & $-$0.16 &  $-$0.35\\
12,816.046  & 20.0    & 3.907 & $-$0.63 &  $-$0.40\\
12,823.868  & 20.0    & 3.907 & $-$0.85 &  $-$0.45\\
11,780.542  & 22.0    & 1.442 & $-$2.17 &  $-$0.55\\
11,797.186  & 22.0    & 1.429 & $-$2.28 &  $-$0.15\\
11,892.768  & 22.0    & 4.175 & $-$2.17 &  $-$0.15\\
11,949.547  & 22.0    & 1.442 & $-$1.57 &  $-$0.55\\
12,569.571  & 22.0    & 2.173 & $-$2.05 &  $-$0.45\\
12,671.092  & 22.0    & 1.429 & $-$2.52 &  $-$0.55\\
12,738.370  & 22.0    & 4.803 & $-$2.35 &  $-$0.25\\
12,738.477  & 22.0    & 4.728 & $-$1.25 &       ---\\
12,811.480  & 22.0    & 2.159 & $-$1.39 &  $-$0.55\\
12,821.672  & 22.0    & 1.459 & $-$1.19 &  $-$0.40\\
12,831.442  & 22.0    & 1.429 & $-$1.49 &  $-$0.55\\
12,840.607  & 22.0    & 4.660 & $-$2.85 &  $-$0.25\\
12,847.033  & 22.0    & 1.442 & $-$1.55 &  $-$0.35\\
11,882.846  & 26.0    & 2.196 & $-$2.17 &  $-$0.51\\
12,190.100  & 26.0    & 3.632 & $-$2.73 &  $-$0.60\\
12,648.943  & 26.0    & 6.395 & $-$2.69 &  $-$0.54\\
\hline\hline
\end{tabular}
\end{table}

\acknowledgments

We thank the staff of La Silla Observatory, European Southern Observatory, in Chile
for their support during our observations.
The development and operation of WINERED have been 
supported by MEXT Programs for the Strategic Research Foundation
at Private Universities (Nos.~S0801061 and S1411028)
and Grants-in-Aid, KAKENHI, from Japan Society for the Promotion of Science
(JSPS; Nos.~16684001, 20340042, 2184005 and 2628028).
We thank the anonymous referee for his/her positive and pertinent suggestions
on the early draft of this letter.


\begin{thebibliography}{}
\expandafter\ifx\csname natexlab\endcsname\relax\def\natexlab#1{#1}\fi
\providecommand{\url}[1]{\href{#1}{#1}}
\providecommand{\dodoi}[1]{doi:~\href{http://doi.org/#1}{\nolinkurl{#1}}}
\providecommand{\doeprint}[1]{\href{http://ascl.net/#1}{\nolinkurl{http://ascl.net/#1}}}
\providecommand{\doarXiv}[1]{\href{https://arxiv.org/abs/#1}{\nolinkurl{https://arxiv.org/abs/#1}}}

\bibitem[{{Alonso} {et~al.}(1999){Alonso}, {Arribas}, \&
  {Mart{\'{\i}}nez-Roger}}]{alonso99}
{Alonso}, A., {Arribas}, S., \& {Mart{\'{\i}}nez-Roger}, C. 1999, \aaps, 140,
  261, \dodoi{10.1051/aas:1999521}

\bibitem[{{Asplund} {et~al.}(2009){Asplund}, {Grevesse}, {Sauval}, \&
  {Scott}}]{asplund09}
{Asplund}, M., {Grevesse}, N., {Sauval}, A.~J., \& {Scott}, P. 2009, \araa, 47,
  481, \dodoi{10.1146/annurev.astro.46.060407.145222}

\bibitem[{{Bessell} {et~al.}(1998){Bessell}, {Castelli}, \& {Plez}}]{bessell98}
{Bessell}, M.~S., {Castelli}, F., \& {Plez}, B. 1998, \aap, 333, 231

\bibitem[{{Bono} {et~al.}(2017){Bono}, {Braga}, {Ferraro}, {Fiorentino},
  {Gilmozzi}, {Iannicola}, {Magurno}, {Matsunaga}, {Monelli}, \&
  {Rastello}}]{bono17}
{Bono}, G., {Braga}, V.~F., {Ferraro}, I., {et~al.} 2017, in IAU Symposium,
  Vol. 316, Formation, Evolution, and Survival of Massive Star Clusters, ed.
  C.~{Charbonnel} \& A.~{Nota}, 36--43

\bibitem[{{Buzzoni} {et~al.}(2010){Buzzoni}, {Patelli}, {Bellazzini}, {Pecci},
  \& {Oliva}}]{buzzoni10}
{Buzzoni}, A., {Patelli}, L., {Bellazzini}, M., {Pecci}, F.~F., \& {Oliva}, E.
  2010, \mnras, 403, 1592, \dodoi{10.1111/j.1365-2966.2009.16223.x}

\bibitem[{{Campbell} {et~al.}(2013){Campbell}, {D'Orazi}, {Yong},
  {Constantino}, {Lattanzio}, {Stancliffe}, {Angelou}, {Wylie-de Boer}, \&
  {Grundahl}}]{campbell13}
{Campbell}, S.~W., {D'Orazi}, V., {Yong}, D., {et~al.} 2013, \nat, 498, 198,
  \dodoi{10.1038/nature12191}

\bibitem[{{Cardelli} {et~al.}(1989){Cardelli}, {Clayton}, \&
  {Mathis}}]{cardelli89}
{Cardelli}, J.~A., {Clayton}, G.~C., \& {Mathis}, J.~S. 1989, \apj, 345, 245,
  \dodoi{10.1086/167900}

\bibitem[{{Carretta} {et~al.}(2009){Carretta}, {Bragaglia}, {Gratton},
  {D'Orazi}, \& {Lucatello}}]{carretta09}
{Carretta}, E., {Bragaglia}, A., {Gratton}, R., {D'Orazi}, V., \& {Lucatello},
  S. 2009, \aap, 508, 695, \dodoi{10.1051/0004-6361/200913003}

\bibitem[{{Chiavassa} {et~al.}(2018){Chiavassa}, {Casagrande}, {Collet},
  {Magic}, {Bigot}, {Thevenin}, \& {Asplund}}]{chiavassa18}
{Chiavassa}, A., {Casagrande}, L., {Collet}, R., {et~al.} 2018, ArXiv e-prints.
\newblock \doarXiv{1801.01895}

\bibitem[{{Clement} {et~al.}(2001){Clement}, {Muzzin}, {Dufton}, {Ponnampalam},
  {Wang}, {Burford}, {Richardson}, {Rosebery}, {Rowe}, \& {Hogg}}]{clement01}
{Clement}, C.~M., {Muzzin}, A., {Dufton}, Q., {et~al.} 2001, \aj, 122, 2587,
  \dodoi{10.1086/323719}

\bibitem[{{Da Silva} {et~al.}(2016){Da Silva}, {Lemasle}, {Bono}, {Genovali},
  {McWilliam}, {Cristallo}, {Bergemann}, {Buonanno}, {Fabrizio}, {Ferraro},
  {Fran{\c c}ois}, {Iannicola}, {Inno}, {Laney}, {Kudritzki}, {Matsunaga},
  {Nonino}, {Primas}, {Przybilla}, {Romaniello}, {Th{\'e}venin}, \&
  {Urbaneja}}]{dasilva16}
{Da Silva}, R., {Lemasle}, B., {Bono}, G., {et~al.} 2016, \aap, 586, A125,
  \dodoi{10.1051/0004-6361/201527300}

\bibitem[{{D'Orazi} {et~al.}(2017){D'Orazi}, {Desidera}, S. and {Gratton}, R.~G. and {Lanza}, A.~F. and 
	{Messina}, S. and {Andrievsky}, S.~M. and {Korotin}, S. and 
	{Benatti}, S. and {Bonnefoy}, M. and {Covino}, E. and {Janson}, M.
	}]{dorazi17}
	{D'Orazi}, V., {Desidera}, S., {Gratton}, R.~G.,  {et al.} 2017, \aap, 598, 19,
	\dodoi{10.1051/0004-6361/201629283}

\bibitem[{{Fiorentino} {et~al.}(2015){Fiorentino}, {Bono}, {Monelli},
  {Stetson}, {Tolstoy}, {Gallart}, {Salaris}, {Mart{\'{\i}}nez-V{\'a}zquez}, \&
  {Bernard}}]{fiorentino15a}
{Fiorentino}, G., {Bono}, G., {Monelli}, M., {et~al.} 2015, \apjl, 798, L12,
  \dodoi{10.1088/2041-8205/798/1/L12}

\bibitem[{{Frogel} \& {Whitelock}(1998)}]{frogel98}
{Frogel}, J.~A., \& {Whitelock}, P.~A. 1998, \aj, 116, 754,
  \dodoi{10.1086/300478}

\bibitem[{{Gonzalez} {et~al.}(2011){Gonzalez}, {Rejkuba}, {Zoccali}, {Hill},
  {Battaglia}, {Babusiaux}, {Minniti}, {Barbuy}, {Alves-Brito}, {Renzini},
  {Gomez}, \& {Ortolani}}]{gonzalez11}
{Gonzalez}, O.~A., {Rejkuba}, M., {Zoccali}, M., {et~al.} 2011, \aap, 530, A54,
  \dodoi{10.1051/0004-6361/201116548}

\bibitem[{{Gratton} {et~al.}(2012){Gratton}, {Carretta}, \&
  {Bragaglia}}]{gratton12}
{Gratton}, R.~G., {Carretta}, E., \& {Bragaglia}, A. 2012, \aapr, 20, 50,
  \dodoi{10.1007/s00159-012-0050-3}

\bibitem[{{Gullikson} {et~al.}(2014){Gullikson}, {Dodson-Robinson}, \&
  {Kraus}}]{gullikson14}
{Gullikson}, K., {Dodson-Robinson}, S., \& {Kraus}, A. 2014, \aj, 148, 53,
  \dodoi{10.1088/0004-6256/148/3/53}

\bibitem[{{Gustafsson} {et~al.}(2008){Gustafsson}, {Edvardsson}, {Eriksson},
  {J{\o}rgensen}, {Nordlund}, \& {Plez}}]{gustafsson08}
{Gustafsson}, B., {Edvardsson}, B., {Eriksson}, K., {et~al.} 2008, \aap, 486,
  951, \dodoi{10.1051/0004-6361:200809724}

\bibitem[{{Harris}(1996)}]{harris96}
{Harris}, W.~E. 1996, \aj, 112, 1487, \dodoi{10.1086/118116}

\bibitem[{{Hron} {et~al.}(2015){Hron}, {Uttenthaler}, {Aringer}, {Klotz},
  {Lebzelter}, {Paladini}, \& {Wiedemann}}]{hron15}
{Hron}, J., {Uttenthaler}, S., {Aringer}, B., {et~al.} 2015, \aap, 584, A27,
  \dodoi{10.1051/0004-6361/201526862}

\bibitem[{{Ikeda} {et~al.}(2016){Ikeda}, {Kobayashi}, {Kondo}, {Otsubo},
  {Hamano}, {Sameshima}, {Yoshikawa}, {Fukue}, {Nakanishi}, {Kawanishi},
  {Nakaoka}, {Kinoshita}, {Kitano}, {Asano}, {Takenaka}, {Watase}, {Mito},
  {Yasui}, {Minami}, {Izumu}, {Yamamoto}, {Mizumoto}, {Arasaki}, {Arai},
  {Matsunaga}, \& {Kawakita}}]{ikeda16}
{Ikeda}, Y., {Kobayashi}, N., {Kondo}, S., {et~al.} 2016, in \procspie, Vol.
  9908, Ground-based and Airborne Instrumentation for Astronomy VI, 99085Z

\bibitem[{{Kochanek} {et~al.}(2017){Kochanek}, {Shappee}, {Stanek}, {Holoien},
  {Thompson}, {Prieto}, {Dong}, {Shields}, {Will}, {Britt}, {Perzanowski}, \&
  {Pojma{\'n}ski}}]{kochanek17}
{Kochanek}, C.~S., {Shappee}, B.~J., {Stanek}, K.~Z., {et~al.} 2017, \pasp,
  129, 104502, \dodoi{10.1088/1538-3873/aa80d9}

\bibitem[{{Kunder} {et~al.}(2013){Kunder}, {Stetson}, {Cassisi}, {Layden},
  {Bono}, {Catelan}, {Walker}, {Paredes Alvarez}, {Clem}, {Matsunaga},
  {Salaris}, {Lee}, \& {Chaboyer}}]{kunder13}
{Kunder}, A., {Stetson}, P.~B., {Cassisi}, S., {et~al.} 2013, \aj, 146, 119,
  \dodoi{10.1088/0004-6256/146/5/119}

\bibitem[{{Lapenna} {et~al.}(2015){Lapenna}, {Mucciarelli}, {Ferraro},
  {Origlia}, {Lanzoni}, {Massari}, \& {Dalessandro}}]{lapenna15}
{Lapenna}, E., {Mucciarelli}, A., {Ferraro}, F.~R., {et~al.} 2015, \apj, 813,
  97, \dodoi{10.1088/0004-637X/813/2/97}

\bibitem[{{Lawler} {et~al.}(2013){Lawler}, {Guzman}, {Wood}, {Sneden}, \&
  {Cowan}}]{lawler13}
{Lawler}, J.~E., {Guzman}, A., {Wood}, M.~P., {Sneden}, C., \& {Cowan}, J.~J.
  2013, \apjs, 205, 11, \dodoi{10.1088/0067-0049/205/2/11}

\bibitem[{{Lebzelter} {et~al.}(2014){Lebzelter}, {Nowotny}, {Hinkle},
  {H{\"o}fner}, \& {Aringer}}]{lebzelter14}
{Lebzelter}, T., {Nowotny}, W., {Hinkle}, K.~H., {H{\"o}fner}, S., \&
  {Aringer}, B. 2014, \aap, 567, A143, \dodoi{10.1051/0004-6361/201424078}

\bibitem[{{Lebzelter} {et~al.}(2015){Lebzelter}, {Nowotny}, {Hinkle},
  {H{\"o}fner}, {Aringer}, \& {Heiter}}]{lebzelter15}
{Lebzelter}, T., {Nowotny}, W., {Hinkle}, K.~H., {et~al.} 2015, in Astronomical
  Society of the Pacific Conference Series, Vol. 497, Why Galaxies Care about
  AGB Stars III: A Closer Look in Space and Time, ed. F.~{Kerschbaum}, R.~F.
  {Wing}, \& J.~{Hron}, 283

\bibitem[{{MacLean} {et~al.}(2016){MacLean}, {Campbell}, {De Silva},
  {Lattanzio}, {D'Orazi}, {Simpson}, \& {Momany}}]{maclean16}
{MacLean}, B.~T., {Campbell}, S.~W., {De Silva}, G.~M., {et~al.} 2016, \mnras,
  460, L69, \dodoi{10.1093/mnrasl/slw073}

\bibitem[{{Mart{\'{\i}}nez-V{\'a}zquez}
  {et~al.}(2016){Mart{\'{\i}}nez-V{\'a}zquez}, {Stetson}, {Monelli}, {Bernard},
  {Fiorentino}, {Gallart}, {Bono}, {Cassisi}, {Dall'Ora}, {Ferraro},
  {Iannicola}, \& {Walker}}]{martinezvazquez16b}
{Mart{\'{\i}}nez-V{\'a}zquez}, C.~E., {Stetson}, P.~B., {Monelli}, M., {et~al.}
  2016, \mnras, 462, 4349, \dodoi{10.1093/mnras/stw1895}

\bibitem[{{Matsunaga}(2006)}]{matsunaga06b}
{Matsunaga}, N. 2006, PhD thesis, The University of Tokyo

\bibitem[{{Matsunaga} {et~al.}(2009){Matsunaga}, {Feast}, \&
  {Menzies}}]{matsunaga09}
{Matsunaga}, N., {Feast}, M.~W., \& {Menzies}, J.~W. 2009, \mnras, 397, 933,
  \dodoi{10.1111/j.1365-2966.2009.14992.x}

\bibitem[{{Mura-Guzm{\'a}n} {et~al.}(2018){Mura-Guzm{\'a}n}, {Villanova},
  {Mu{\~n}oz}, \& {Tang}}]{mura18}
{Mura-Guzm{\'a}n}, A., {Villanova}, S., {Mu{\~n}oz}, C., \& {Tang}, B. 2018,
  \mnras, 474, 4541, \dodoi{10.1093/mnras/stx2918}

\bibitem[{{Nowotny} {et~al.}(2010){Nowotny}, {H{\"o}fner}, \&
  {Aringer}}]{nowotny10}
{Nowotny}, W., {H{\"o}fner}, S., \& {Aringer}, B. 2010, \aap, 514, A35,
  \dodoi{10.1051/0004-6361/200911899}

\bibitem[{{Origlia} {et~al.}(2013){Origlia}, {Oliva}, {Maiolino},
  {Mucciarelli}, {Baffa}, {Biliotti}, {Bruno}, {Falcini}, {Gavriousev},
  {Ghinassi}, {Giani}, {Gonzalez}, {Leone}, {Lodi}, {Massi}, {Montegriffo},
  {Mochi}, {Pedani}, {Rossetti}, {Scuderi}, {Sozzi}, \& {Tozzi}}]{origlia13}
{Origlia}, L., {Oliva}, E., {Maiolino}, R., {et~al.} 2013, \aap, 560, A46,
  \dodoi{10.1051/0004-6361/201322586}

\bibitem[{{Otsubo} {et~al.}(2016){Otsubo}, {Ikeda}, {Kobayashi}, {Sukegawa},
  {Kondo}, {Hamano}, {Sameshima}, {Fukue}, {Yoshikawa}, {Nakanishi}, {Watase},
  {Takenaka}, {Asano}, {Yasui}, {Matsunaga}, \& {Kawakita}}]{otsubo16}
{Otsubo}, S., {Ikeda}, Y., {Kobayashi}, N., {et~al.} 2016, in \procspie, Vol.
  9908, Ground-based and Airborne Instrumentation for Astronomy VI, 990879

\bibitem[{{Pancino} {et~al.}(2010){Pancino}, {Rejkuba}, {Zoccali}, \&
  {Carrera}}]{pancino10}
{Pancino}, E., {Rejkuba}, M., {Zoccali}, M., \& {Carrera}, R. 2010, \aap, 524,
  A44, \dodoi{10.1051/0004-6361/201014383}

\bibitem[{{Pancino} {et~al.}(2017){Pancino}, {Romano}, {Tang}, {Tautvai{\v
  s}ien{\.e}}, {Casey}, {Gruyters}, {Geisler}, {San Roman}, {Randich},
  {Alfaro}, {Bragaglia}, {Flaccomio}, {Korn}, {Recio-Blanco}, {Smiljanic},
  {Carraro}, {Bayo}, {Costado}, {Damiani}, {Jofr{\'e}}, {Lardo}, {de Laverny},
  {Monaco}, {Morbidelli}, {Sbordone}, {Sousa}, \& {Villanova}}]{pancino17}
{Pancino}, E., {Romano}, D., {Tang}, B., {et~al.} 2017, \aap, 601, A112,
  \dodoi{10.1051/0004-6361/201730474}

\bibitem[{{Pritzl} {et~al.}(2005){Pritzl}, {Venn}, \& {Irwin}}]{pritzl05b}
{Pritzl}, B.~J., {Venn}, K.~A., \& {Irwin}, M. 2005, \aj, 130, 2140,
  \dodoi{10.1086/432911}

\bibitem[{{Ram{\'{\i}}rez} \& {Allende Prieto}(2011)}]{ramirez11}
{Ram{\'{\i}}rez}, I., \& {Allende Prieto}, C. 2011, \apj, 743, 135,
  \dodoi{10.1088/0004-637X/743/2/135}

\bibitem[{{Sameshima}(2018)}]{sameshima18}
{Sameshima}, H. 2018, Submittetd to PASP


\bibitem[{{Shappee} {et~al.}(2014){Shappee}, {Prieto}, {Grupe}, {Kochanek},
  {Stanek}, {De Rosa}, {Mathur}, {Zu}, {Peterson}, {Pogge}, {Komossa}, {Im},
  {Jencson}, {Holoien}, {Basu}, {Beacom}, {Szczygie{\l}}, {Brimacombe},
  {Adams}, {Campillay}, {Choi}, {Contreras}, {Dietrich}, {Dubberley},
  {Elphick}, {Foale}, {Giustini}, {Gonzalez}, {Hawkins}, {Howell}, {Hsiao},
  {Koss}, {Leighly}, {Morrell}, {Mudd}, {Mullins}, {Nugent}, {Parrent},
  {Phillips}, {Pojmanski}, {Rosing}, {Ross}, {Sand}, {Terndrup}, {Valenti},
  {Walker}, \& {Yoon}}]{shappee14}
{Shappee}, B.~J., {Prieto}, J.~L., {Grupe}, D., {et~al.} 2014, \apj, 788, 48,
  \dodoi{10.1088/0004-637X/788/1/48}

\bibitem[{{Simmerer} {et~al.}(2013){Simmerer}, {Feltzing}, \&
  {Primas}}]{simmerer13b}
{Simmerer}, J., {Feltzing}, S., \& {Primas}, F. 2013, \aap, 556, A58,
  \dodoi{10.1051/0004-6361/201220723}

\bibitem[{{Sloan} {et~al.}(2010){Sloan}, {Matsunaga}, {Matsuura}, {Zijlstra},
  {Kraemer}, {Wood}, {Nieusma}, {Bernard-Salas}, {Devost}, \&
  {Houck}}]{sloan10}
{Sloan}, G.~C., {Matsunaga}, N., {Matsuura}, M., {et~al.} 2010, \apj, 719,
  1274, \dodoi{10.1088/0004-637X/719/2/1274}

\bibitem[{{Smith} {et~al.}(2006){Smith}, {Price}, \& {Moffett}}]{smith06}
{Smith}, B.~J., {Price}, S.~D., \& {Moffett}, A.~J. 2006, \aj, 131, 612,
  \dodoi{10.1086/497972}

\bibitem[{{Sneden}(1973)}]{sneden73}
{Sneden}, C.~A. 1973, PhD thesis, The University of Texas at Austin

\bibitem[{{Taniguchi} {et~al.}(2018){Taniguchi}, {Matsunaga}, {Kobayashi},
  {Fukue}, {Hamano}, {Ikeda}, {Kawakita}, {Kondo}, {Sameshima}, \&
  {Yasui}}]{taniguchi18}
{Taniguchi}, D., {Matsunaga}, N., {Kobayashi}, N., {et~al.} 2018, \mnras, 473,
  4993, \dodoi{10.1093/mnras/stx2691}

\bibitem[{{Uttenthaler} {et~al.}(2015){Uttenthaler}, {Blommaert}, {Wood},
  {Lebzelter}, {Aringer}, {Schultheis}, \& {Ryde}}]{uttenthaler15}
{Uttenthaler}, S., {Blommaert}, J.~A.~D.~L., {Wood}, P.~R., {et~al.} 2015,
  \mnras, 451, 1750, \dodoi{10.1093/mnras/stv1052}

\bibitem[{{Wang} {et~al.}(2016){Wang}, {Primas}, {Charbonnel}, {Van der
  Swaelmen}, {Bono}, {Chantereau}, \& {Zhao}}]{wang16}
{Wang}, Y., {Primas}, F., {Charbonnel}, C., {et~al.} 2016, \aap, 592, A66,
  \dodoi{10.1051/0004-6361/201628502}

\bibitem[{{Wood}(1979)}]{wood79}
{Wood}, P.~R. 1979, \apj, 227, 220, \dodoi{10.1086/156721}

\bibitem[{{Zoccali} \& {Valenti}(2016)}]{zoccali16}
{Zoccali}, M., \& {Valenti}, E. 2016, \pasa, 33, e025,
  \dodoi{10.1017/pasa.2015.56}

\end{thebibliography}
\end{document}